\documentclass[sigconf,authordraft, review=false]{acmart}

\usepackage{amsmath}
\usepackage{mathtools}
\usepackage{amsthm}
\usepackage{amsxtra}
\usepackage{amsfonts}
\usepackage{graphicx}
\usepackage{latexsym}
\usepackage{epsfig}
\usepackage{verbatim}
\usepackage{bm}
\usepackage{bbm}
\usepackage{color}
\usepackage{varwidth}
\usepackage{subfigure}
\usepackage{wrapfig}
\usepackage{trivfloat}
\usepackage{todonotes}
\usepackage{url}
\usepackage{fullpage}
\usepackage{dsfont}
\usepackage{balance}
\usepackage{fullpage}
\usepackage{algorithm}
\usepackage{algpseudocode}
\usepackage{times}
\usepackage{fancyhdr,graphicx,amsmath}

\usepackage{hyperref}

\usepackage{caption}

\usepackage{epstopdf}
\PrependGraphicsExtensions{.tif, .tiff}

\AtBeginDocument{%
  \providecommand\BibTeX{{%
    \normalfont B\kern-0.5em{\scshape i\kern-0.25em b}\kern-0.8em\TeX}}}

\begin{document}

\title{A Pre-training Oracle for Predicting Distances in Social Networks}



\author{Gunjan Mahindre}
\affiliation{%
  \institution{Electrical and Computer Engineering, Colorado State University}
  \city{Fort Collins}
  \country{USA}
}
\author{Randy Paffenroth}
\affiliation{%
  \institution{Mathematical Sciences, Worcester Polytechnic Institute}
  \city{Worcester}
  \country{USA}
}
\author{Anura Jayasumana}
\affiliation{%
  \institution{Electrical and Computer Engineering, Colorado State University}
  \city{Fort Collins}
  \country{USA}
}
\author{Rasika Karkare}
\affiliation{%
  \institution{Data Science, Worcester Polytechnic Institute}
  \city{Worcester}
  \country{USA}
}







\begin{abstract}
  
In this paper, we propose a novel method to make distance predictions in real-world social networks.
As predicting missing distances is a difficult problem, we take a two-stage approach. 
Structural parameters for families of synthetic networks are first estimated from a small set of measurements of a real-world network
and these synthetic networks are then used to pre-train the predictive neural networks.
Since our model first searches for the most suitable synthetic graph parameters which can be used as an ``oracle'' to create arbitrarily large training data sets, we call our approach ``Oracle Search Pre-training” (OSP).

For example, many real-world networks exhibit a Power law structure in their node degree distribution, so a Power law model can provide a foundation for the desired oracle to generate synthetic pre-training networks, if the appropriate Power law graph parameters can be estimated. 
Accordingly, we conduct experiments on real-world Facebook, Email, and Train Bombing networks and show that OSP outperforms models without pre-training, models pre-trained with inaccurate parameters, and other distance prediction schemes such as Low-rank Matrix Completion. \emph{In particular, we achieve a prediction error of less than one hop with only 1\% of sampled distances from the social network.}
OSP can be easily extended to other domains such as random networks by choosing an appropriate model to generate synthetic training data, and therefore promises to impact many different network learning problems.

\end{abstract}





\maketitle

\section{Introduction}

Shortest node-pair distances in graphs pack a lot of insight regarding the network as a whole and thus have been used in several graph analysis algorithms. In social network analysis, for example,  hop-distances are used for predicting friendships \cite{link1, link2}, 
extracting sensor network topology \cite{ton, tpm}, and modeling social networks \cite{model1, model2} to name a few.
However, 
missing measurements in sampled datasets adversely affect the structural properties of social networks \cite{missingdata, missingdata2} and the original network characteristics are misrepresented. Predicting these missing measurements would solve this problem. 

Neural networks have proved to be useful in solving many prediction problems but they often require a large amount of training data to perform well. Pre-training has alleviated this issue with transfer learning, especially when the second task is much more difficult for training \cite{pretrain1, pretrain2}.
However, relatively little is known about the pre-training behavior related to neural networks used for social network analysis as social networks are difficult to measure or sample completely due to their cost of measurement, privacy, or storage issues. The ever increasing size of social networks today only adds to this problem.

We demonstrate a prediction model based on efficient pre-training for distance inference in social networks when only sparse samples are known. We use artificially simulated training data to compensate for the lack of sufficient real-world measurements. For accurate prediction, this artificial pre-training data has to be faithful with the characteristics of the target social network.
However, we do not know the characteristics of the target network. All we have is a small fraction of randomly measured node-pair distances. A way to estimate the right pre-training parameters is required.
The central idea of this paper is to build an ``Oracle'' that points us towards these suitable set of pre-training parameters. 

\begin{figure}[htp!]
\vspace{-2mm}
\centering
\includegraphics[width=\linewidth]{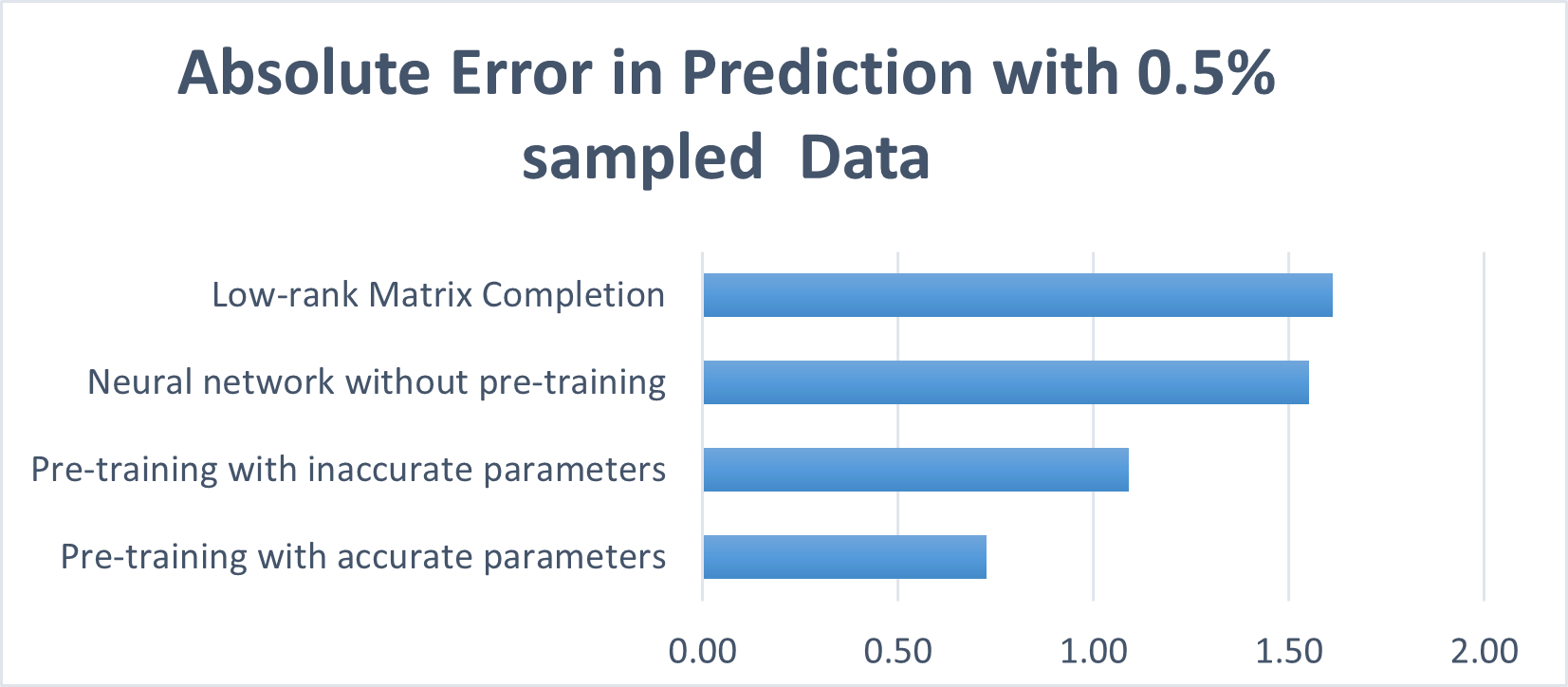}
\caption{Comparison of proposed method with Low-rank Matrix Completion, neural network without pre-training, and pre-training with inaccurate parameters on a real world Virgili Emails network with only 0.5\% of distances sampled. OSP improves the performance by 53\% from a neural network without pre-training and by 33\% from when inaccurate pre-training parameters are used.}
\label{fig:comparison}
\vspace{-2mm}
\end{figure}

Fig. \ref{fig:comparison} gives a preview of our results for a 1133 node real-world Email network when only 0.5\% of the total node-pair distances are known. We can see that accurate pre-training parameters improve the prediction performance by 33\% over pre-training with inaccurate parameters and by 53\% over models without pre-training.

We list the main contributions of our work in Section \ref{sec:contribution}. Section \ref{sec:related work} discusses relevant literature. 
Required theoretical 
background is provided in Section \ref{sec:theory} with the complete methodology in Section \ref{sec:methodology}. Sections \ref{sec:results} and \ref{sec:discussion} provide performance analysis and sensitivity study of the model. 
We summarize our work in Section \ref{sec:conclusion}.



\section{Contribution}
\label{sec:contribution}

The main objective of this paper is to present a sophisticated method of efficiently pre-training an autoencoder (AE) for prediction when sufficient target measurements are not available for training.
A multi-stage setup for estimating network characteristics from partial distance measurements is proposed.
We integrate the idea of pre-training with that of best parameter selection for optimal prediction.

These optimally pre-trained autoencoders are demonstrated to predict missing shortest node-pair distances in social networks.
We show the effectiveness of OSP on real-world social networks such as Virgili Emails, Facebook, and Train Bombing network which vary in their size and areas of function.
The prediction performance is also compared against an existing distance inference model based on Low-rank Matrix Completion \cite{MC_candes, ton, LCN2019} (Section \ref{sec:results}).

We improve with respect to the existing work in the following ways:

\begin{enumerate}
    \item Artificially generated training data is used to pre-train an autoencoder when sufficient real-world data is not available (Section \ref{data generation}). 
    \item A novel two stage oracle is presented to predict suitable parameters for optimal pre-training. The oracle points us towards the suitable parameters solely using the sampled distances and without knowing any characteristics of the target network (Section \ref{model}).
    \item OSP does not require large amounts of measurements from the target network that needs to be predicted. Our experiments show results for networks sampled as low as 0.1\%, 0.5\%, and 1\% while also covering up to 80\% sampling (Section \ref{sec:results}).
    \item Sensitivity of the predictive model is studied towards different pre-training parameters. This helps us understand pre-training behaviour of social networks and identify best set of pre-training parameters for the given test network parameters (Section \ref{sec:discussion}).
\end{enumerate}


\section{Related Work}
\label{sec:related work}

In this section, we discuss the most relevant literature to this paper: prediction techniques for node-pair distances, completing missing network measurements from partial measurements, and using pre-training for predicting distances.

Low-rank Matrix Completion (LMC) \cite{MC_candes} has been used effectively in reconstructing missing distances in graphs such as IoT or social networks when partial distance measurements are known, either randomly or from a selected set of landmarks. It has demonstrated accurate prediction up to 20-40\% of sampled measurements \cite{ton, landmarks, LCN2019}. This deterministic technique works well due to the low-rankness of the distance matrices. Thus we use Low-rank Matrix Completion as one of the reference points for comparing our results.

Deep learning techniques have found its way into social network analysis for distance prediction.
Graph convolutional neural networks have been used to predict missing links in recommender systems \cite{graph_MC}.
Preserved distances of each node from a few landmark nodes are successfully used \cite{VC_MLP} to train a multi-layer perceptron (MLP) and predict distance between two vertices without a high space cost. 
WiDE, \cite{wifi}, uses unsupervised stacked autoencoders to pre-train a deep neural network 
and predicts person-to-person distances using surrounding Wifi signals.
However, these techniques rely on a significant amount of measurements within the real-world networks. 

Our work is inspired by the methods proposed in \cite{LCN2020, blind} where Hadamard Autoencoders are trained on synthetic data when only sparse measurements are available from the target network. Though this helps us train on artificial data, all social networks are not the same and vary greatly in their characteristics. This leaves room for improving the pre-training process to be customized for each social network being reconstructed. 

Thus, we hypothesize that building a pre-training ``Oracle'' that guides us into a narrower range of the pre-training data being used will help significantly improve the prediction performance by generating customized training data that is more faithful to the given social network.


\section{Theory}
\label{sec:theory}

In this section, we provide necessary background by giving a brief introduction of the autoencoders and the data used. We also briefly talk about the concept of pre-training which forms the foundation of this work.

\subsection{Supervised Autoencoders}

An autoencoder is a special type of artificial neural network that tries to reproduce the input at the output while learning its representation in a low dimensional space.  
It consists of two parts, namely, encoder and decoder. The encoder takes the input, $\textbf{x}$, and maps it to a hidden representation $\textbf{h}$, such that:

\begin{equation}
    \textbf{h} = \sigma(\textbf{W}\textbf{x} + \textbf{b}),
    \label{eq:hidden representation}
\end{equation}

\noindent where $\textbf{W}$ is the weight matrix, $\textbf{b}$ is the bias vector and $\sigma$ is the activation function for the encoder. We use Leaky ReLU as the activation function throughout the autoencoder. It is defined as:

\begin{equation}
    \sigma(z) = 
    \begin{cases}
        z, & \textit{if } z>0\\
        \alpha z, & z<=0,
    \end{cases}
\end{equation}
\label{eq:relu}

\noindent where $z$ is the input and $\alpha$ is a small non-zero, constant gradient. Leaky ReLU helps to solve the ``vanishing gradient'' problem caused by ReLU where the function outputs 0 if $z<0$, seizing the update of the hyperparameters and thus the learning.

The decoder reconstructs this hidden representation $\textbf{h}$ to the output, $\hat{\textbf{x}}$, given by:

\begin{equation}
    \hat{\textbf{x}} = \sigma (\textbf{W}'\textbf{h} + \textbf{b}'),
    \label{eq:output representation}
\end{equation}

\noindent where $\textbf{W}'$ is the weight matrix, $\textbf{b}'$ is the bias vector and $\sigma$ is the activation function for the decoder.

The input and output layers have the same number of neurons in order to reconstruct the given input. Each neuron operates on the incoming signal and generates an output based on its activation function, weight and bias values. The weight and bias values are found by training so that the difference between $\hat{\textbf{x}}$ and $\textbf{x}$ is minimized.


\subsection{Training Protocols}

We use supervised training method to train the hyperparameters, i.e., while training, the expected output is known. Back-propagation is used to train the weights and biases for each layer during training. Back-propagation consists of two phases. During the first phase, input is fed to the autoencoder with weights and biases initialized to random values and the output layer reconstructs the input. In the second phase, the error between expected and predicted output is calculated using a mean squared error loss function given by Eq. \ref{eq:loss function}. The weights and biases are updated iteratively using this error value.

\begin{equation}
    \mathcal{L}(\textbf{x},\hat{\textbf{x}}) =  || \textbf{x} - \hat{\textbf{x}} || ^2,
\label{eq:loss function}
\end{equation}

\noindent where $\textbf{x}$ is the known expected output, $\hat{\textbf{x}}$ is the predicted output as reconstructed by the output layer.
Thus from Eq. (\ref{eq:hidden representation}), (\ref{eq:output representation}), and (\ref{eq:loss function}), we can say that

\begin{equation}
    \mathcal{L}(\textbf{x},\hat{\textbf{x}}) = ||\textbf{x} - \sigma (\textbf{W}'(\sigma(\textbf{W}\textbf{x} + \textbf{b})) + \textbf{b}')|| ^2 .
    \label{eq:loss function 2}
\end{equation}

In other words, the loss function is the mean $(\frac{1}{n} \sum\limits_{i=1}^{n})$ of the squared errors $( {x}_i - \hat{x}_i )^2$, which is an easily computable quantity for the given data sample, and can be expressed as $ MSE =  \frac{1}{n} \sum\limits_{i=1}^{n} ( {x}_i - \hat{{x}}_i )^2$,
\noindent where $n$ is the total number of samples. 


The size of the hidden layer can be adjusted in proportion to the size of the target data. We used learning rate 0.001, batch size 1, and hidden layer size 50 for the Facebook network and 20 for the Virgili Emails and Train Bombing network, as determined from the observed measurements.

\subsection{Pre-training}
\label{pretraining}

The concept of pre-training is inspired from the way humans learn. While learning new things, we re-use our knowledge of the past.
Pre-training a neural network refers to training the model during one task and using the trained parameters as a head start for another task. This way, the model does not have to learn from scratch for the second task. It is especially useful if the two tasks are similar or the final task is much more 
difficult and can be broken down in simpler pre-training tasks \cite{pretrain1, pretrain2} .
For instance, training an autoencoder to predict missing measurements of sparsely available data can be difficult. Instead, we pre-train the autoencoder on readily available simulated data and try to predict the parameters for this pre-training data so that it is similar in characteristics to the target data.

Analysis of both natural and human-created real-world networks have shown that such networks closely follow specific rules, such as Power law behaviour in their degree distributions $P(k) \sim k^{-\gamma}$, with the exponent varying between 2 and 3. The characteristic that only a few nodes have an extremely large number of connections while most of the nodes have very few links is due to `preferential attachment' \cite{albert2002statistical, yule1925ii, barabasi2016network}. Furthermore, social networks like Facebook and WeChat are observed to have Power law growth dynamics \cite{8280512}.
Thus, for a given real-world network, it is possible to identify a closely related network type that can be used to generate synthetic data for training. Social networks that we are interested in for this paper, e.g., Emails, Facebook, and Train Bombing networks, are well documented to follow Power law.

Training the autoencoder on Power law networks with similar parameters helps the model treat the target real-world network as just a variation of the training network. Another advantage is that we can generate sufficient data as required for proper training and effective performance of the neural network.

\subsection{Data}
Given a graph $G=\left\{V, E\right\}$, where $V$ is the set of nodes in the network and $E$ is the set of undirected edges, we use shortest hop-distances between node-pairs as our data. For a given graph $G$, its \emph{shortest hop-distance matrix} $\textbf{H}$ can be given as
\begin{equation}
\textbf{H} = \left[h_{ji}=h_{ij} = \textit{shortest hop-distance from }i \textit{ to } j \right],
\label{H definition}
\end{equation}
\noindent 
The shortest \emph{Hop-Distance} Matrix (HDM),  $\textbf{H}
\in \mathbb{N}^{N \times N}_0$ can represent the entire network,
where
$ \mathbb{N}_0$ denotes the non-negative integers $\mathbb{N}\cup\{0\} $ and $N$ denotes number of nodes in $G$. 

Note that nodes and links may have different interpretation in different networks, e.g., in the Facebook network, nodes are users and links represent friendships, where as in a software system network, nodes represent software modules and links may represent dependency between the two modules.
While our work focuses only on \emph{connected, unweighted, and undirected graphs}, there are several commercially important social networks in this domain such as Friendship, Email, crime, and interaction networks to name a few.



\subsection{Sampling Scheme}
We aim to predict missing distances in networks when the target network is only partially sampled. 
\begin{figure}[hb]
\vspace{-4mm}
\centering
\includegraphics[width=2.0in]{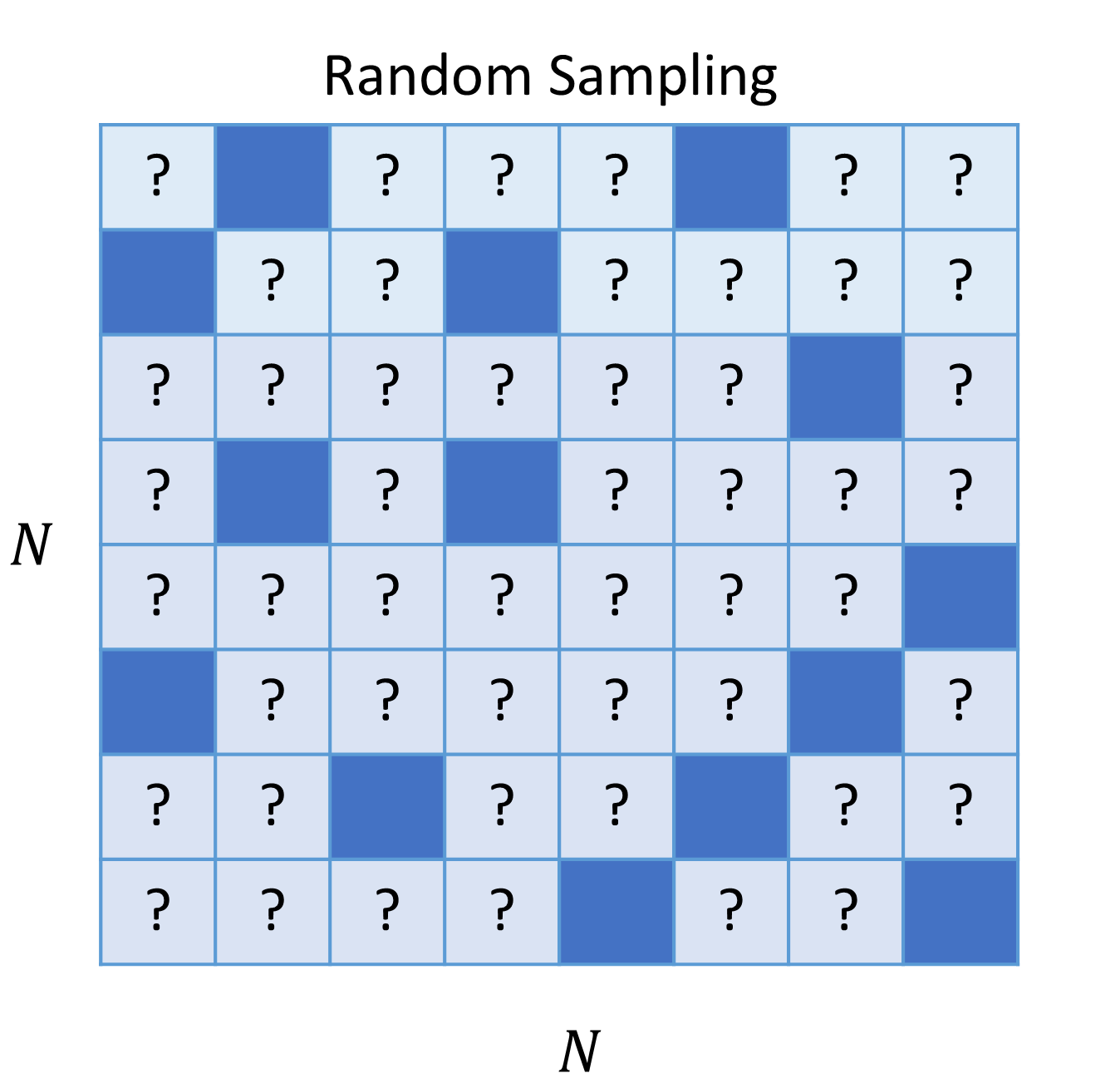}
\vspace*{-3mm}
\caption{Visual illustrations of $\textbf{P}$ (sampled $\textbf{H}$) for random sampling scheme with $N = 8$. Known entries are denoted by dark blue and unknown values by light blue.}
\label{fig:sampling_schemes}
\vspace{-3mm}
\end{figure}
In our experiments, we 
measure a small set of distances between random pairs of nodes from the 
distance matrix $\textbf{H}$ to achieve distributed, unbiased sampling. 
The randomly sampled $\textbf{H}$ results in an $N\times N$ matrix $\textbf{P}$, as shown in Fig. \ref{fig:sampling_schemes}, where the sampled entries are depicted by solid squares and `?'s represent missing entries. 
Other sampling schemes such as random walk or snow-ball sampling tend to be biased towards higher degree nodes, producing a misrepresentation of typical large scale, real-world, unstructured networks with non-uniform degree distributions, e.g., Power law degree distribution \cite{random_sampling1}. 


The conventional approach for network reconstruction from a sparse set of random samples are based on Low-rank Matrix Completion which requires a sampling scheme to measure at least one non-zero distance value in each row for the test network \cite{ton}. However, our approach overcomes this restriction.

\subsection{Low-rankness of Data}
Low-rankness in a dataset implies redundancy which allows reconstruction of missing measurements from a small set of entries, i.e., partial measurements. We leverage this while reconstructing distance matrices. 

\begin{figure}[ht]
\vspace{-2mm}
\centering
\includegraphics[width=0.40\textwidth]{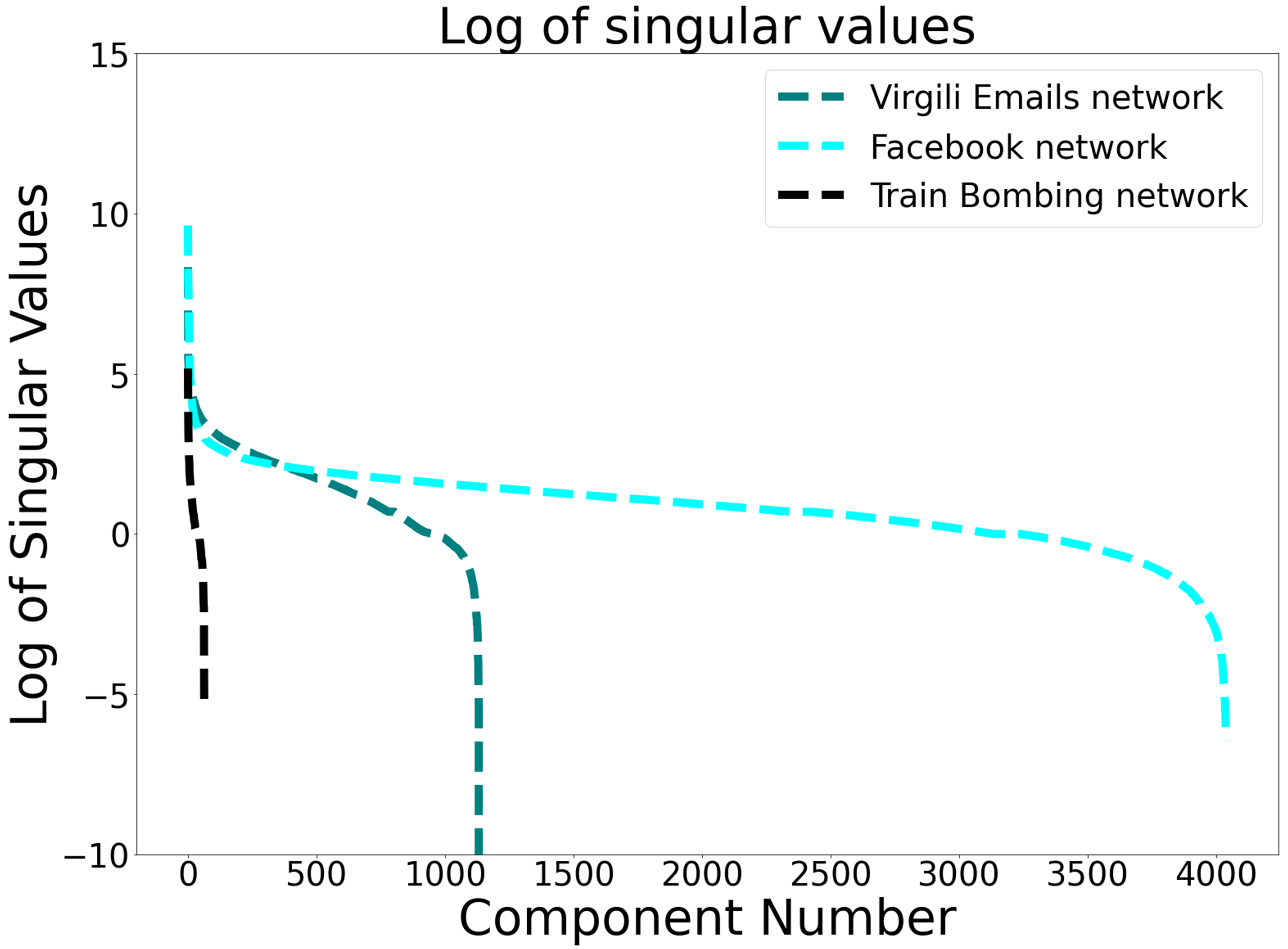}
\vspace*{-3mm}
\caption{Low-rankness of hop-distance matrices of social networks used in this paper.
}
\label{fig:sv}
\vspace{-4mm}
\end{figure}

Here, 
we demonstrate low-rankness of the real-world social networks used.
Fig. \ref{fig:sv} shows the logarithm of all the singular values of $\textbf{H}$. Only a small set of singular values are dominant with the later magnitudes falling down rapidly towards `0', indicating low-rankness of the $\textbf{H}$ matrices for shown networks. 

Note that low-rank data lies on low-dimensional linear subspace whereas low-dimensional data lies on low-dimensional non-linear subspace. 

\section{Methodology}
\label{sec:methodology}

 
 
\begin{figure*}[ht]
\vspace{-1mm}
\centering
\includegraphics[width=\textwidth]{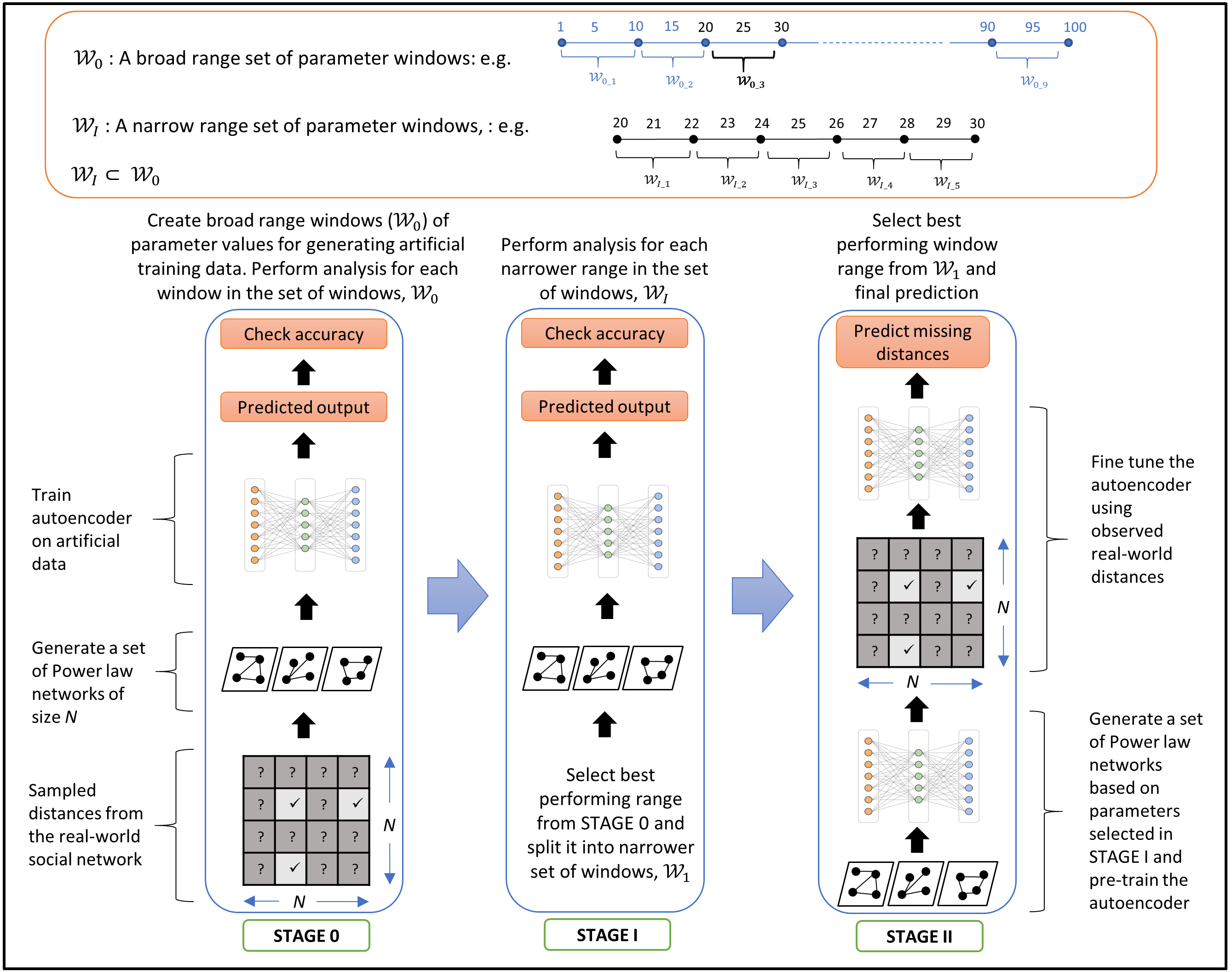}
\vspace*{-5mm}
\caption{Oracle Search Pre-training (OSP): Model architecture. STAGE 0 helps narrow down suitable parameter values from a very broad range and STAGE I helps to select the three most suitable parameters for the given sampled network. The neural network is pre-trained on artificial data generated using these selected parameters and fine tuned using sampled distances form the target network in STAGE II to predict missing distances in the real-world social network.}
\label{fig:flowchart}
\vspace{-3mm}
\end{figure*}

An overview of the proposed simple approach for predicting distances from a small set of measurements is shown in Fig. \ref{fig:flowchart}. 
In this section, we discuss how pre-training data is artificially generated to train a neural network, which parameters are tuned to make the synthetic data similar to real-world data we are predicting on, and how the ideal values of this parameter are estimated using our proposed pre-training ``Oracle''.

\begin{figure*}[ht]
\centering
\includegraphics[width=0.80\textwidth]{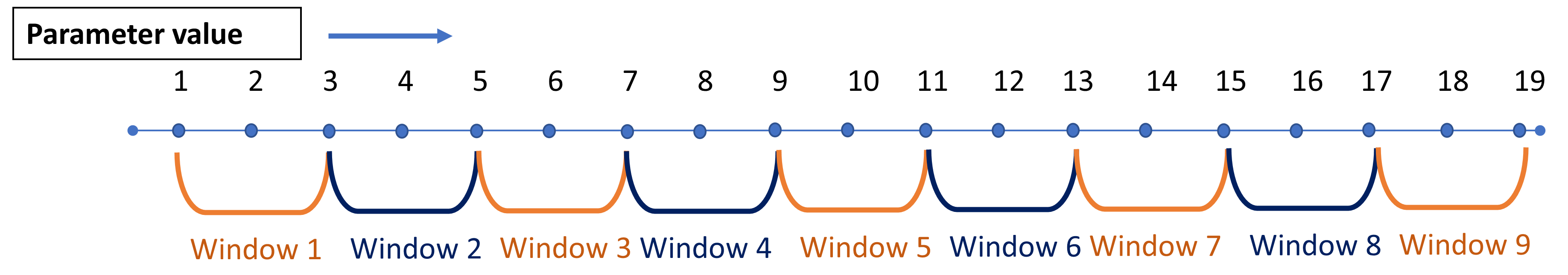}
\vspace*{-3mm}
\caption{Illustration of selecting windows for parameter values for generating artificial Power law networks. Neural network is trained and tested to evaluate effectiveness of each window on prediction.}
\label{fig:windows}
\end{figure*}

\subsection{Generation of Artificial Training Data}
\label{data generation}


We learnt in Section \ref{pretraining} that our target networks, i.e., social networks, are Power law in nature. Thus, we use synthetic Power law networks as a base for generating our pre-training data.

Power law cluster model, in Python's \textbf{\textit{Networkx}} library, creates graphs for a given network size and edges using preferential attachment with the function \textit{\textbf{powerlaw\_cluster\_graph(}N, m, p\textbf{)}}, where a graph of $N$ nodes is grown by attaching new nodes, each with $m$ edges that are preferentially attached to existing nodes with high degree \cite{powerlaw_structure, networkx}, and $p$ is the probability of forming a triangle after adding a random edge. The model generates a network whose degree distribution follows a Power law curve and thus, represents social networks such as protein-protein interactions, World-Wide-Web, citations network, Facebook, and others.
However, depending on the given parameter values,
the generated graphs could be vastly different from the specific network under consideration. Even for two networks of the same type, e.g., social networks, the parameters could be significantly different. Next we address the method for determining the appropriate parameters that need to be tuned, corresponding to the specific network of interest.


\subsection{Tuning Parameter for Synthetic Data Generation}



Node degree is the simplest way of describing how nodes are connected to each other. Yet, it has a prodigious effect on the network characteristics such as clustering coefficient, betweenness centrality, diameter of the network, etc., proving to be a simple but significant network feature.

As $m$ represents number of edges connected to each new node, in the function \textit{\textbf{powerlaw\_cluster\_graph(}N, m, p\textbf{)}}, it is analogous to the degree of the node.
Hence, we focus on finding a suitable range of values for parameter $m$ for the given target network. We generate various training networks with fixed $N$ - to generate networks of the size same as that of the target network, consciously selected values of $m$ - to study the effect of different $m$ values on prediction performance, while using widely spread values of probability $p$, i.e., from 0.1 to 0.9 - to generate networks with different extent of clustering. This allows us to train the autoencoder on different clustering densities, as we do not know the clustering coefficient of our target network.

Next, we use window method, as described below, to systematically vary $m$ and study its effect on prediction performance.

\subsection{Window Method for Varying the Parameter Value}
\label{window method}

Our goal is to find a narrow range of values for $m$ to generate training data so that it comes closer to the characteristics of our target network.

If we assume that the average node degree of the target network is less than some value $n_d$ and that the network has a Power law node degree distribution, then the distribution is highly right-skewed and we can safely say that the set of suitable $m$ values, for training data generation, are more likely to fall on the left side of average node degree as most of the nodes have node degree less than the average node degree of the network. We also demonstrate this in Section \ref{sec:discussion}. 

As $m > 0$ for connected networks, 
we partition the range 1 to $n_d$ in small windows
as shown in Fig. \ref{fig:windows}, where $n_d = 19$ for illustration purposes. An autoencoder is trained on networks generated for the $m$ values in each window and is tested.
This allows us to evaluate effectiveness of each parameter window with respect to its prediction performance. The goal is to find an ideal range of parameters that will give us the best prediction for the given target network.

Note that the assumed value of $n_d$ may vary as per the type of the target network. For example, we know that in a road network, a node (an intersection) is less likely to have more than 10 edges (roads) so $n_d$ can be set to a lower value. \textbf{Setting a higher $n_d$ has no major drawbacks in the model as it only produces a few more windows for analysis but underestimating $n_d$ might make us miss the ideal set of values of} $m$ \textbf{for the target data.}

\subsection{Evaluation Parameters}

We now describe the performance metric used to evaluate prediction performance of our model. We aim to calculate the cumulative error in prediction.
Thus, we measure and compare mean error and absolute hop-distance error to evaluate prediction performance and to demonstrate effectiveness of our model. Note that we measure these error values for the set of node-pairs under test. This can be a set of observed node-pairs or unobserved node pairs depending on the stage of the model and will be mentioned accordingly.

\subsubsection{Mean error}
\label{mean}

We define mean error $M_e$ as follows:

\begin{equation}
M_e =   { \left[ \sum\limits_{\forall \ i,j \in \check{\Omega}} {| \check{h_{ij}}(f)- h_{ij} |} \right]} \bigg/
{\left[ \sum\limits_{\forall \ i,j \in \check{\Omega}} {h_{ij}} \right]},
\label{equation mean error}
\end{equation}

\noindent where, $\check{\Omega}$ is a set of all node pairs under test, $ \check{h_{ij}}(f)$ refers to the predicted shortest hop-distance from node $i$
to node $j$ when $f$ percentage of random measurements are sampled and $h_{ij}$ refers to its original expected value. 
Mean error gives a percentage value of the prediction error with respect to sum of the original hop-distances in the network.


\subsubsection{Absolute hop-distance error}
\label{AHDE}

Absolute hop-distance error (AHDE) is defined as
\begin{equation}
 H_e = 
\bigg\{\sum\limits_{\forall \ i,j \in \check{\Omega}}  | \check{h_{ij}}(f)- h_{ij}|\bigg\}\bigg/\{card(\check{h})\},
\label{hop-error}
\end{equation}
where 
$card(\check{h})$ gives the total number of node-pair entries in $\textbf{H}$ under test. 
AHDE captures average deviation in prediction in the magnitude of hop-distances. For instance, the absolute hop-distance error of 1 implies that on average a predicted path length will be off by 1 hop, i.e., for the original hop-distance of 10, predicted value will be somewhere from 9 to 11.

\subsection{Model}
\label{model}

A visual illustration of the model architecture can be seen in Fig. \ref{fig:flowchart}. 
A detailed explanation of various stages of the proposed model follows where we explain the ideal parameter selection (STAGE 0 and STAGE I), systematic pre-training and final prediction of missing node-pair distances (STAGE II).


\subsubsection{STAGE 0-Broad range parameter selection}


In initial stages of building the Oracle, the autoencoder is evaluated for a range of $m$ values as we are not aware of the ideal values for the tunable parameter. These values are chosen using a sliding window to include a wide range (Section \ref{window method} and Fig. \ref{fig:windows}). As we saw in Section \ref{window method}, a parameter range can be assumed for $m$ to start. If the range is broad, it can be divided into smaller windows. This can be viewed as STAGE 0.




For example, being a communication network, Virgili Emails network might have a high average node degree. So we first test the autoencoder performance on a set of windows $\mathcal{W}_0$ of $m$ values of [1,5,10], [10,15,20], and so on up to [90,95,100] (as shown in Figs. \ref{fig:virgili}[a]).
The best performing window of [1,5,10], say $\mathcal{W}_{0\_k}$, is selected from this set for the next stage, i.e., STAGE I.

Note that STAGE 0 is optional and can be used when a quick narrow down of a possible wide range of $m$ values is required.

\subsubsection{STAGE I-Narrow range parameter selection}

The best performing window from STAGE 0 is further split into a narrower set of windows, $\mathcal{W}_I$ and $\mathcal{W}_I \subset \mathcal{W}_0$ (as shown in Fig. \ref{fig:flowchart}), to tune the parameter values further. The autoencoder is trained for each of these windows, $\mathcal{W}_{I\_k}$, independently and is tested after each session on observed distances from the target network. 

\begin{figure}[ht]
\vspace{-4mm}
\centering
\subfigure[STAGE 0 evaluation]
        {
		\includegraphics[width=0.23\textwidth]{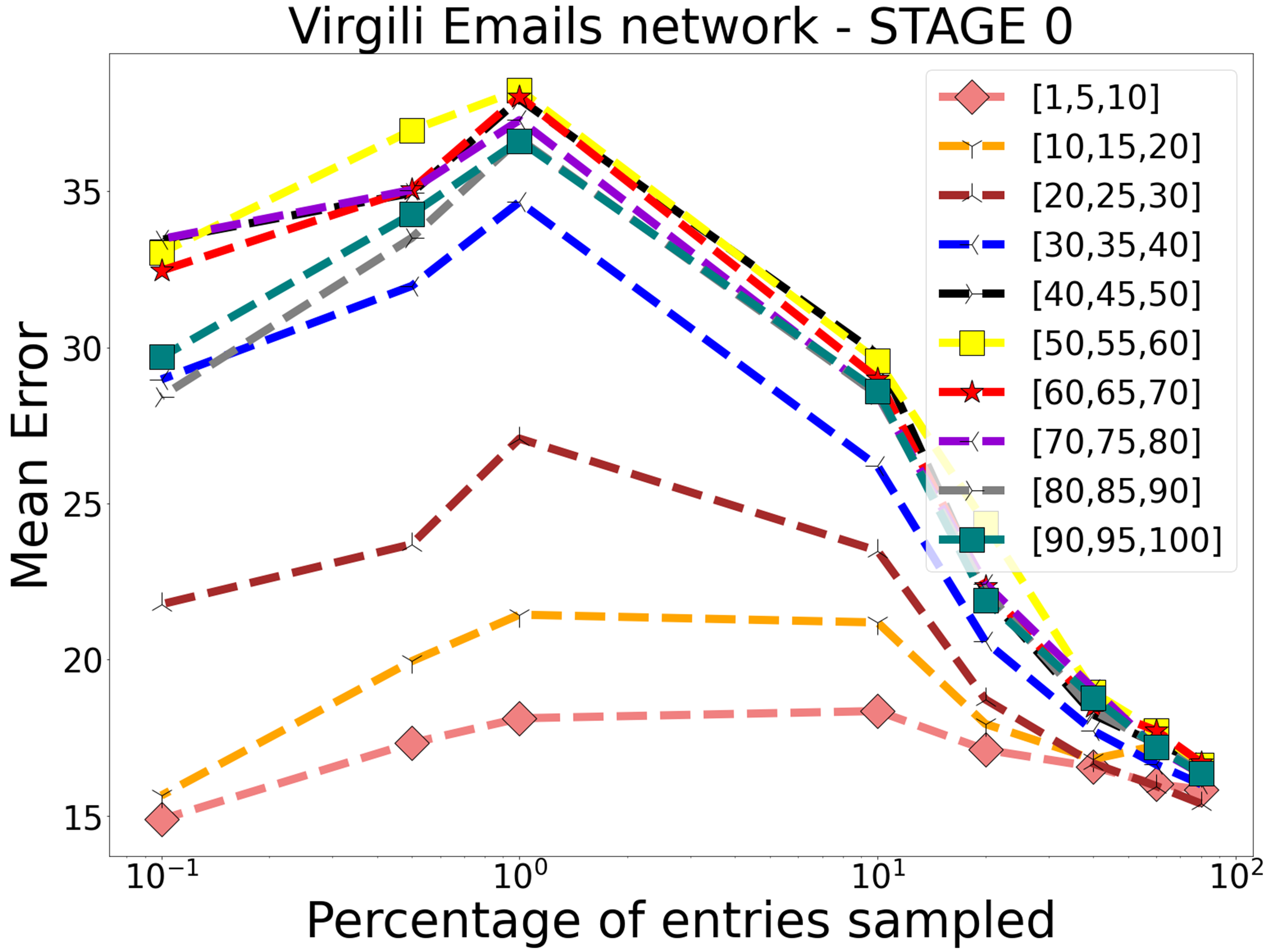}
		 }
\subfigure[STAGE I evaluation]
        {
		\includegraphics[width=0.23\textwidth]{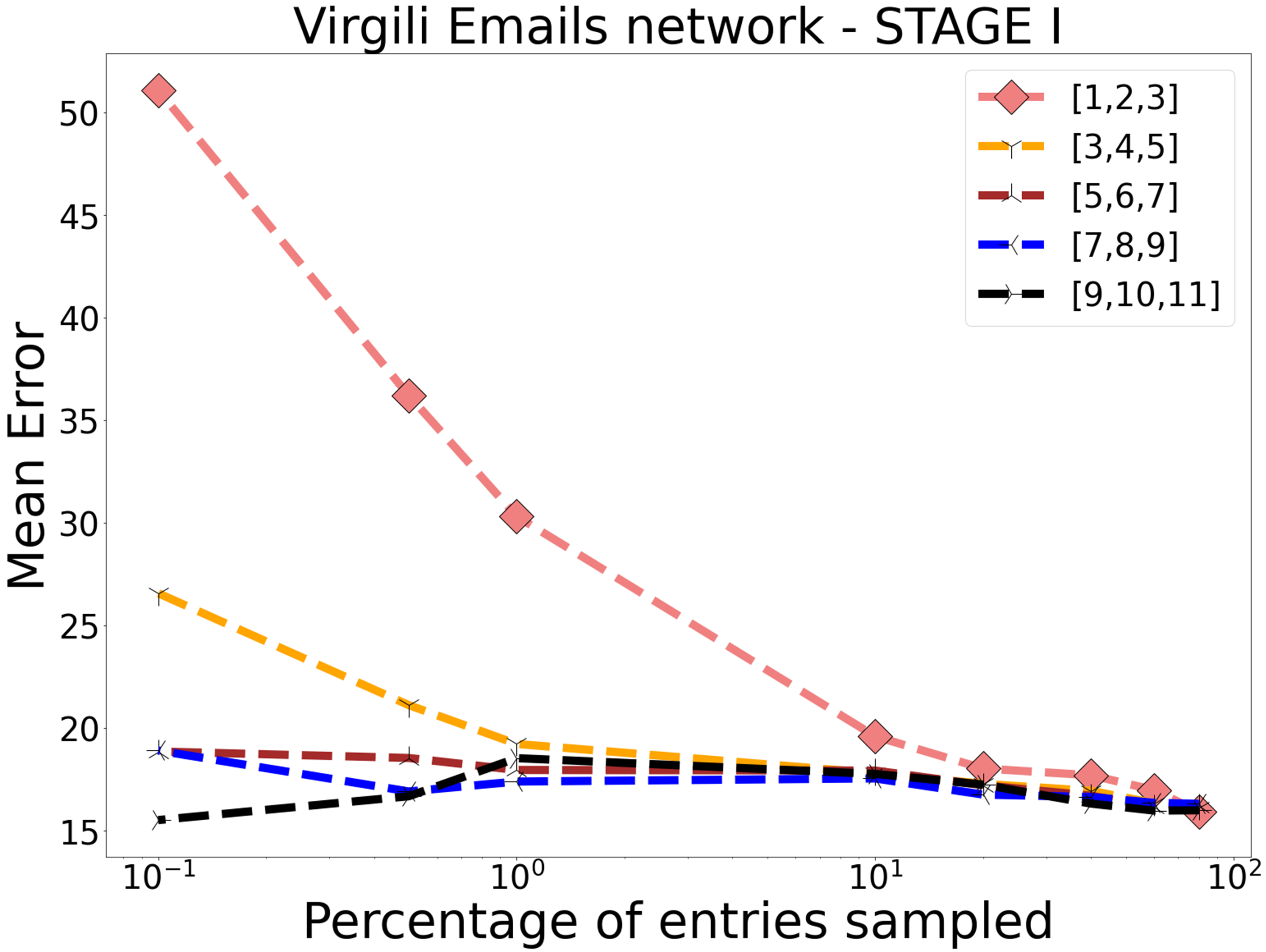}
		} 
\caption{Parameter selection for pre-training in the Oracle for the Virgili Emails network: (a) STAGE 0: Each broad range window is evaluated for its prediction effectiveness on sampled distances. The best parameter range is selected for the next stage. Here, parameter range [1,5,10] is seen to perform the best.
(b) STAGE I: The braod range of (1,5,10) selected from STAGE 0 is split into narrower windows and each window is evaluated. 
}
\label{fig:virgili}
\vspace{-3mm}
\end{figure}

For example, we selected the range [1,5,10] from STAGE 0 for the Virgili Emails network, thus, in STAGE I, we have trained the autoencoder on narrower windows of $m$ values as shown in Fig. \ref{fig:virgili}[b] from [1,2,3], [3,4,5], and so on up to [9,10,11].


\captionsetup[figure]{skip=1pt}


As we are evaluating the system over partial and presumably a very small quantity of sampled data, and because no single window performed significantly better than the rest in STAGE I, we select the top three parameter windows that performed well for the sampled measurements. You can see from Figs. \ref{fig:v_stage2} and \ref{fig:fb_stage2} that windows [5,6,7], [7,8,9], and [9,10,11] perform the best for the Virgili Emails and Facebook network. The Train Bombing network being small in size, we can safely pick a narrower range and train the final model on the best performing window from STAGE 1.

\subsubsection{STAGE II-Final Prediction}
The best parameter values selected from STAGE I are eventually used to generate pre-training data.
Once trained, this model can be fine tuned on observed distances from the target network. We examine the results in both cases (with and without fine tuning) in Section \ref{sec:results}.
The trained autoencoder then reconstructs the complete distance matrix from only a fraction of sampled distances from the test network.

Note that STAGE 0 and STAGE I act as an ``Oracle'' to search the ideal pre-training parameters corresponding to the target network while STAGE II pre-trains AE on artificial data generated with these parameters.

\subsection{Implementation Details}
We used TensorFlow, Google's Machine Learning Library, to build the autoencoder and wrote the script in Python 3.6. We ran the experiments on Google Colab.

Data used and the code for our model are available on Github: https://github.com/anonymous/anonymous


\section{Performance evaluation of the proposed method}
\label{sec:results}

We measure mean error (Section \ref{mean}) and absolute hop-distance error (Section \ref{AHDE}) to evaluate the prediction performance of our model and compare it with the Low-rank Matrix Completion \cite{ton} based approach. Here, the errors are measured only over the unobserved node-pair distances. We know that all diagonal values in a distance matrix are `0' and no off-diagonal entries are `0', so we set all diagonal values of $\hat{\textbf{P}}$ to 0 and round-up all off-diagonal values between 0 and 1 to 1.
\begin{figure*}[htbp!]
\centering
\subfigure[Mean error for prediction in Virgili Emails Network]
        {
		\includegraphics[width=0.45\textwidth]{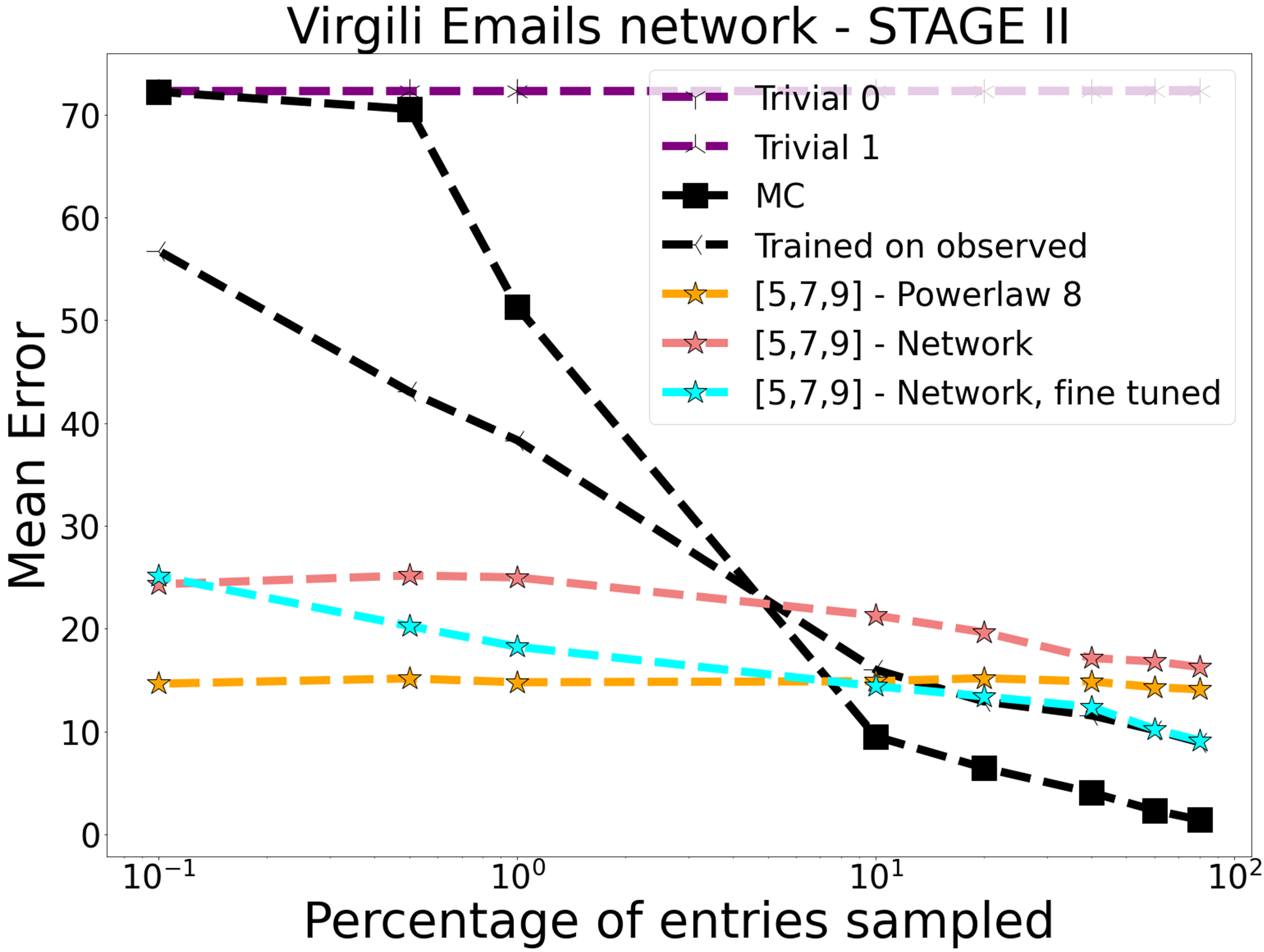}
		 }
\subfigure[AHDE for prediction in Virgili Emails Network]
        {
		\includegraphics[width=0.45\textwidth]{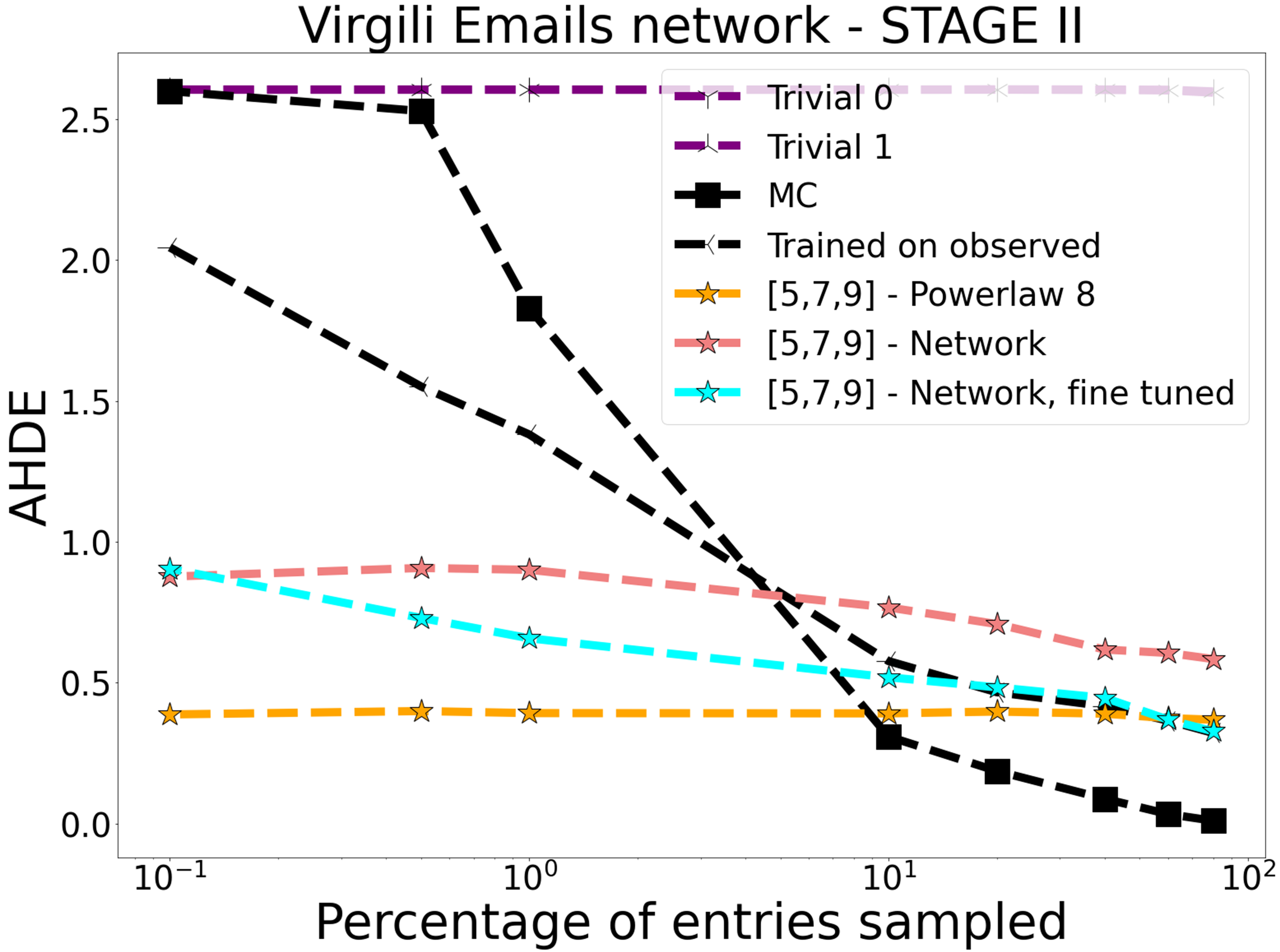}
		} 
\caption{Virgili Emails network: Prediction performance. The plots show that pre-trained neural networks (yellow, pink and blue) outperform prediction especially when lower percentages of entries are sampled from the real-world network. Graceful degradation is also observed for pre-training models. Fine tuning along with pre-training predicts with higher accuracy than a neural network trained only on sampled entries.}
\label{fig:v_stage2}
\end{figure*}

\begin{figure*}[htbp!]
\centering
\subfigure[Mean error for prediction in Facebook Network]
        {
		\includegraphics[width=0.45\textwidth]{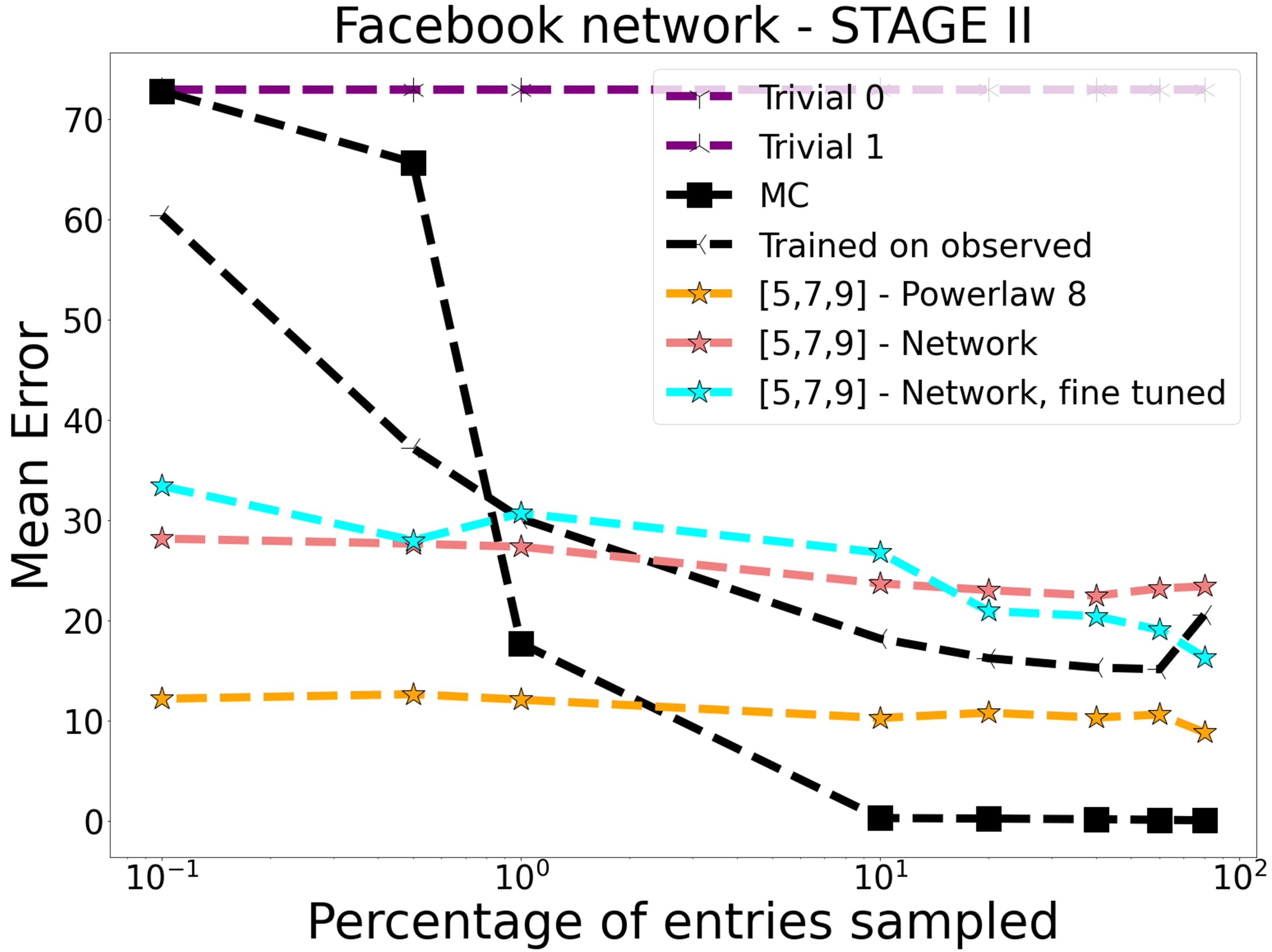}
		 }
\subfigure[AHDE for prediction in Facebook Network]
        {
		\includegraphics[width=0.45\textwidth]{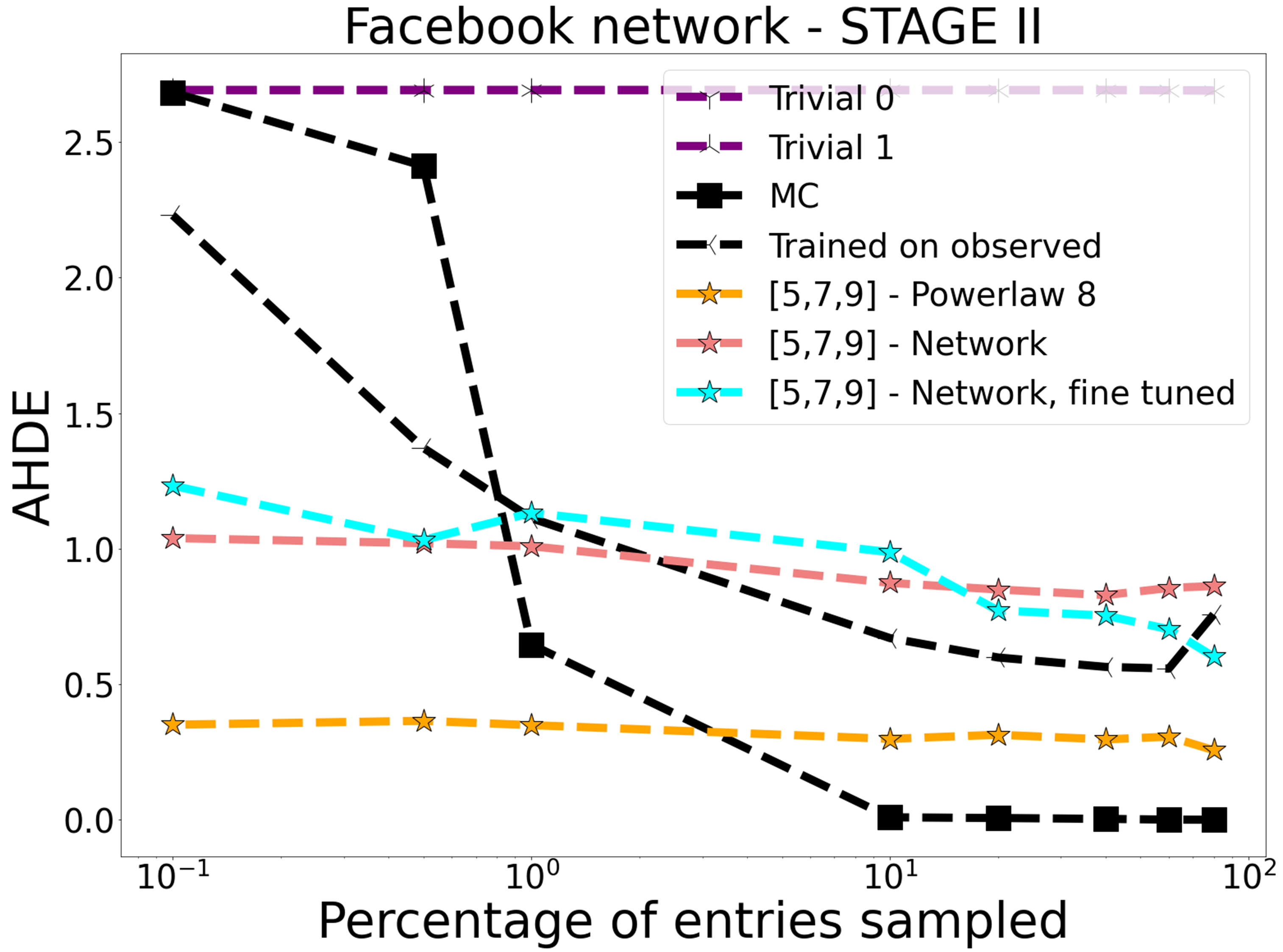}
		} 
\caption{Facebook network: Prediction performance. The plots show that pre-trained neural networks (yellow, pink and blue) outperform prediction especially when lower percentages of entries are sampled from the real-world network. Graceful degradation is also observed for pre-training models. Fine tuning along with pre-training predicts with higher accuracy than a neural network trained only on sampled entries.}
\label{fig:fb_stage2}
\end{figure*}

\begin{figure*}[htbp!]
\centering
\subfigure[Mean error for prediction in Train Bombing Network]
        {
		\includegraphics[width=0.45\textwidth]{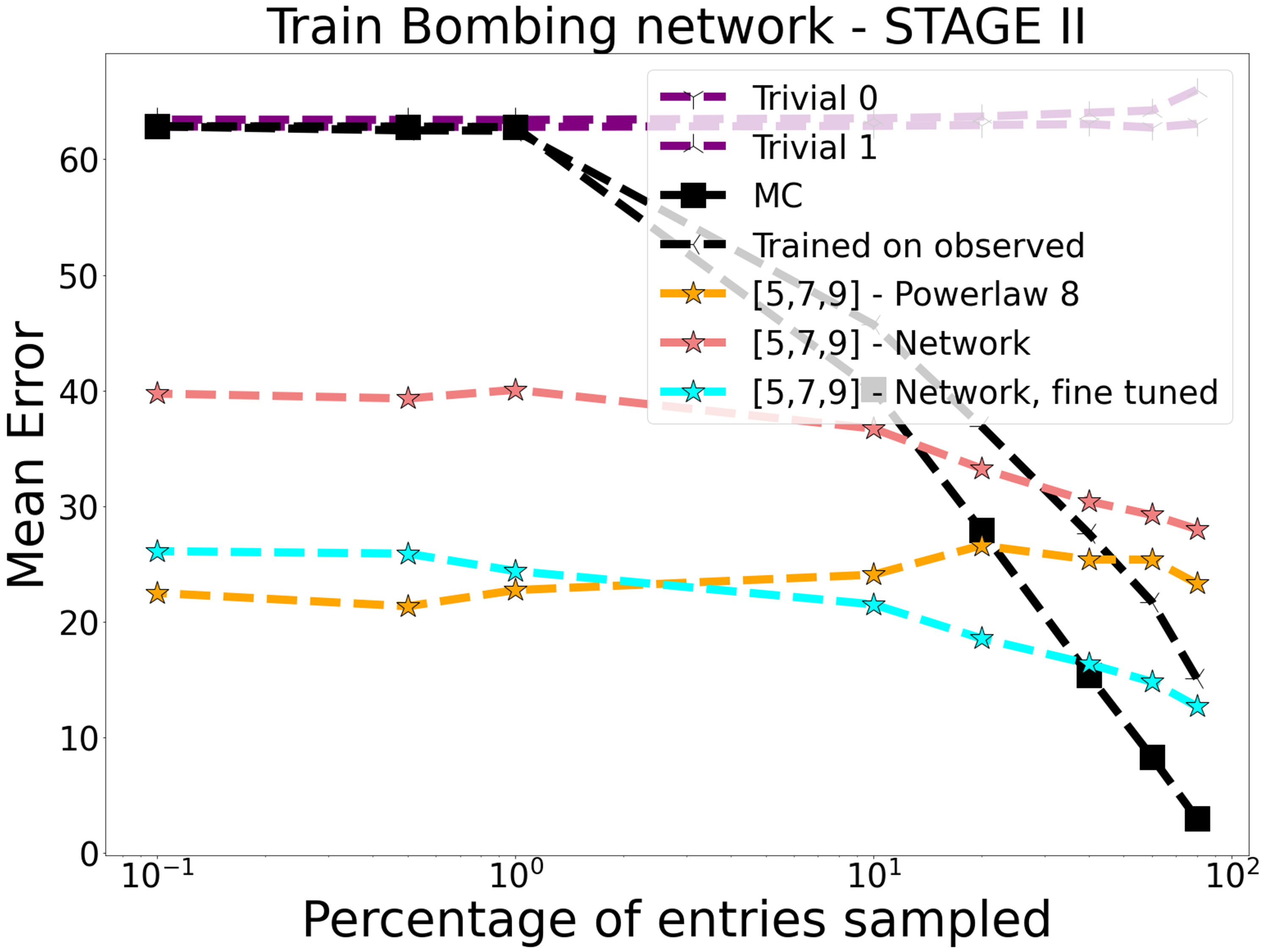}
		 }
\subfigure[AHDE for prediction in Train Bombing Network]
        {
		\includegraphics[width=0.45\textwidth]{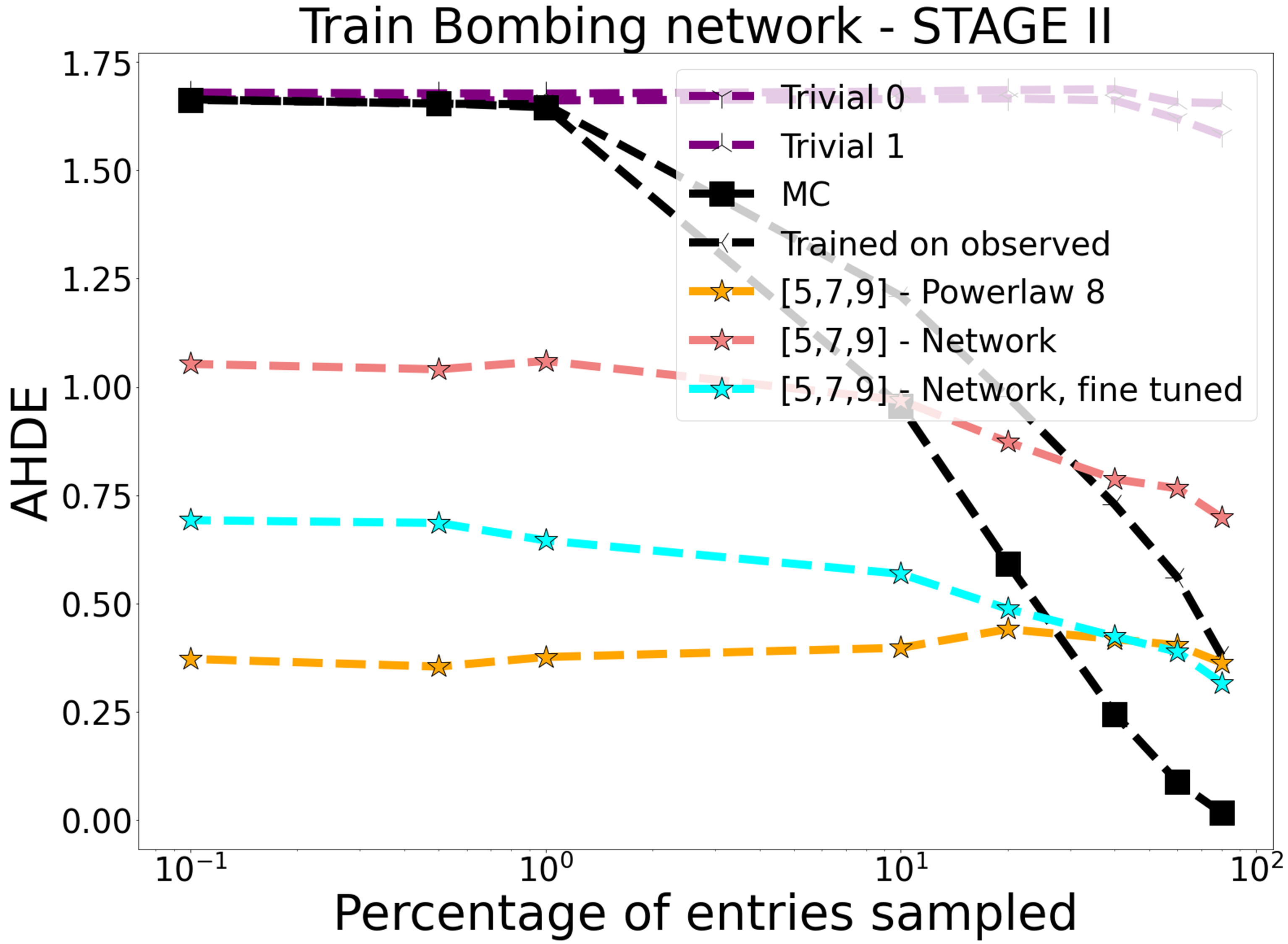}
		} 
\caption{Train Bombing network: Prediction performance. The plots show that pre-trained neural networks (yellow, pink and blue) outperform prediction especially when lower percentages of entries are sampled from the real-world network. Graceful degradation is also observed for pre-training models. Fine tuning along with pre-training predicts with higher accuracy than a neural network trained only on sampled entries.}
\label{fig:tb_stage2}
\vspace{-2mm}
\end{figure*}


The model has been evaluated for different variations and are labeled as shown in Figs. \ref{fig:v_stage2}, \ref{fig:fb_stage2}, and \ref{fig:tb_stage2}. These labels are explained below: 


\begin{itemize}
    \item \textbf{Trivial 0}: all missing values are replaced by 0
    \item \textbf{Trivial 1}: all missing values are replaced by 1 \\
    Trivial 0 and Trivial 1 cases provide a reference to the worse case model performance.
    \item \textbf{MC}: prediction using Low-rank Matrix Completion
    \item \textbf{Trained on observed}: the AE has been trained only on the sampled node-pair distances from the real-world network and no pre-training has been done at all
    \item \textbf{[x,y,z] - Powerlaw q}: the STAGE II AE, i.e., trained on artificial Power law networks with parameter $m$ values x,y, and z, was tested on a same size Power law network artificially generated with $m=q$
    \item \textbf{[x,y,z] - Network}: the STAGE II AE, i.e., trained on artificial Power law networks with parameter $m$ values x,y, and z, was tested on the real-world social network
    \item \textbf{[x,y,z] - Network, fine tuned}: the STAGE II AE, i.e., trained on artificial Power law networks with parameter $m$ values x,y, and z, was fine tuned with the observed entries of the real-world network and then tested on the unobserved entries of the same real-world network
\end{itemize}

Now let us take a look at the performance of the model for three real-world social networks.

\textbf{Virgili Emails network}: This is a 1133 nodes network with average node degree of 9.6. 

\textbf{Facebook network}: This is a 4039 node network with average node degree of 43.6.

\textbf{Train Bombing network}: This is a 64 node network with average node degree of 7.5.

From STAGE 0 and STAGE I of the Virgili Emails network, as shown in Fig. \ref{fig:virgili}, three best windows were selected, [5,6,7], [7,8,9], and [9,10,11]. In STAGE II, the AE is thus trained on [5,7,9] to cover these three windows. This behavior is also seen in the Facebook network. 
The STAGE 0 and STAGE I results for the Facebook and Train Bombing networks are not shown here due to space restrictions. However, can be found at Github: https://github.com/anonymous/anonymous



\textbf{Note that the parameter values selected by the ``Oracle'' are not the same as the average node degree of the test network but tend to be on the lower side of the average node degree value.} We will explore this relation and the sensitivity of the model towards parameter value $m$ in Section \ref{sec:discussion}.

The STAGE II  (Virgili Emails: Fig. \ref{fig:v_stage2}, Facebook: Fig. \ref{fig:fb_stage2}, Train Bombing: Fig. \ref{fig:tb_stage2}) show that AE performs far better than MC for lower percentages of sampled measurements. We also see that AE trained on artificial data performs better than AE trained only on the observed entries as the neural network has much more data to train on from synthetic Power law networks. Note that the network performs best for a variation of Power law network (see label ``[x,y,z]-Powerlaw'') which is obvious as the network was trained on very similar data. The error, when tested on the real-world network is slightly higher than for ``[x,y,z]-Powerlaw'' as the real-world network is different than the synthetic Power law networks. However, we notice that fine tuning the AE almost always bridges this gap. Especially when we have a higher percentage of sampled node-pair measurements from the network, fine tuning gives lower error than mere Power law trained AE for all three networks.

It is worthwhile to say that OSP performs well not only for large networks but also for smaller networks such as the Train Bombing network here, showing independence of the model performance with respect to the network size.

\section{Discussion on optimum parameter selection for pre-training}
\label{sec:discussion}

This section furthers the intuition behind parameter selection for optimal pre-training when the average node degree of the test network is (tentatively) known. 

\begin{figure}[htb!]
\vspace{-1mm}
\centering
\subfigure[Prediction performance with various window sizes.]
        {
		\includegraphics[width=0.20\textwidth]{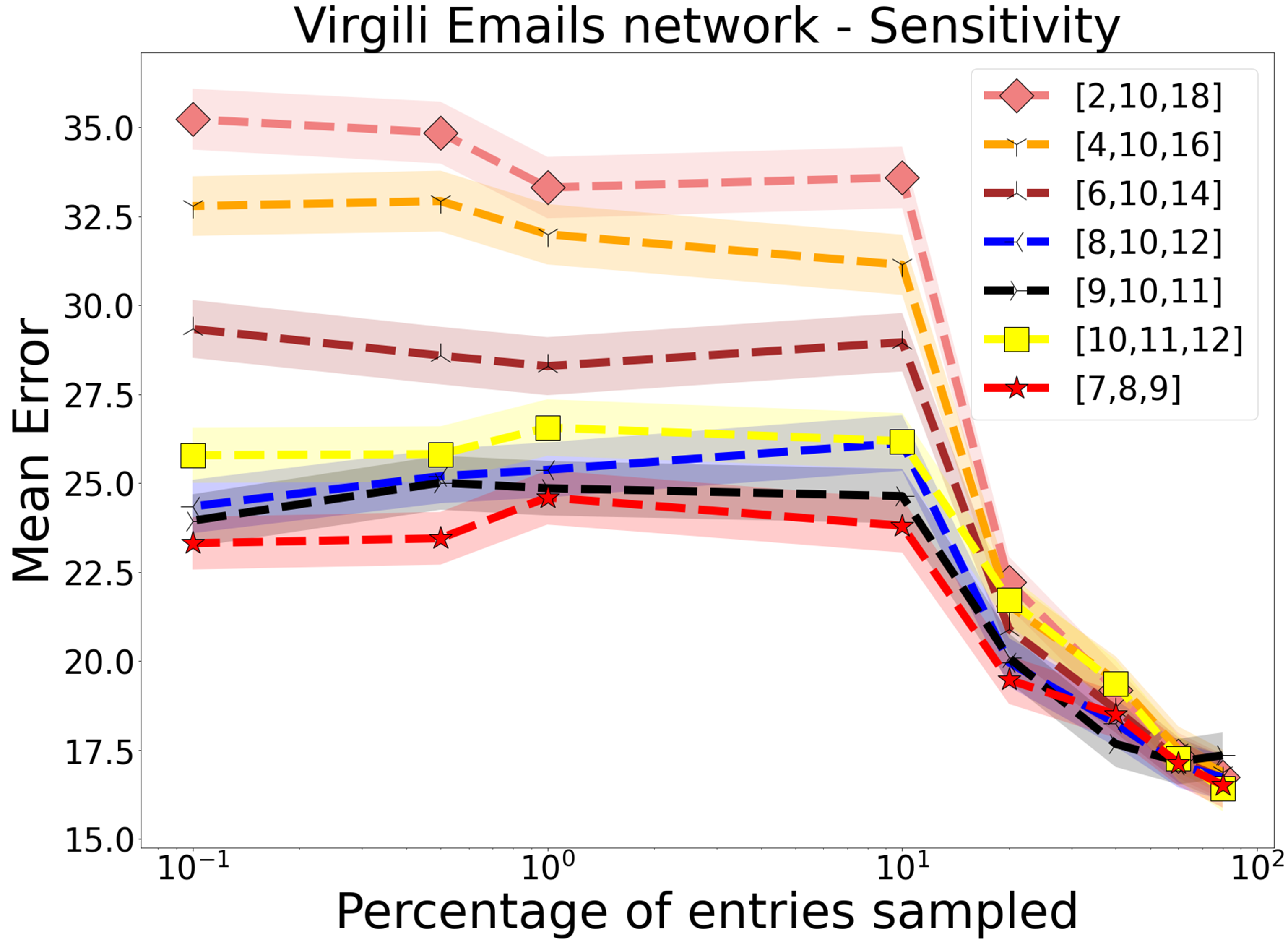}
		}
\subfigure[Prediction performance with various window sizes: Visual comparison.]
        {
		\includegraphics[width=0.24\textwidth]{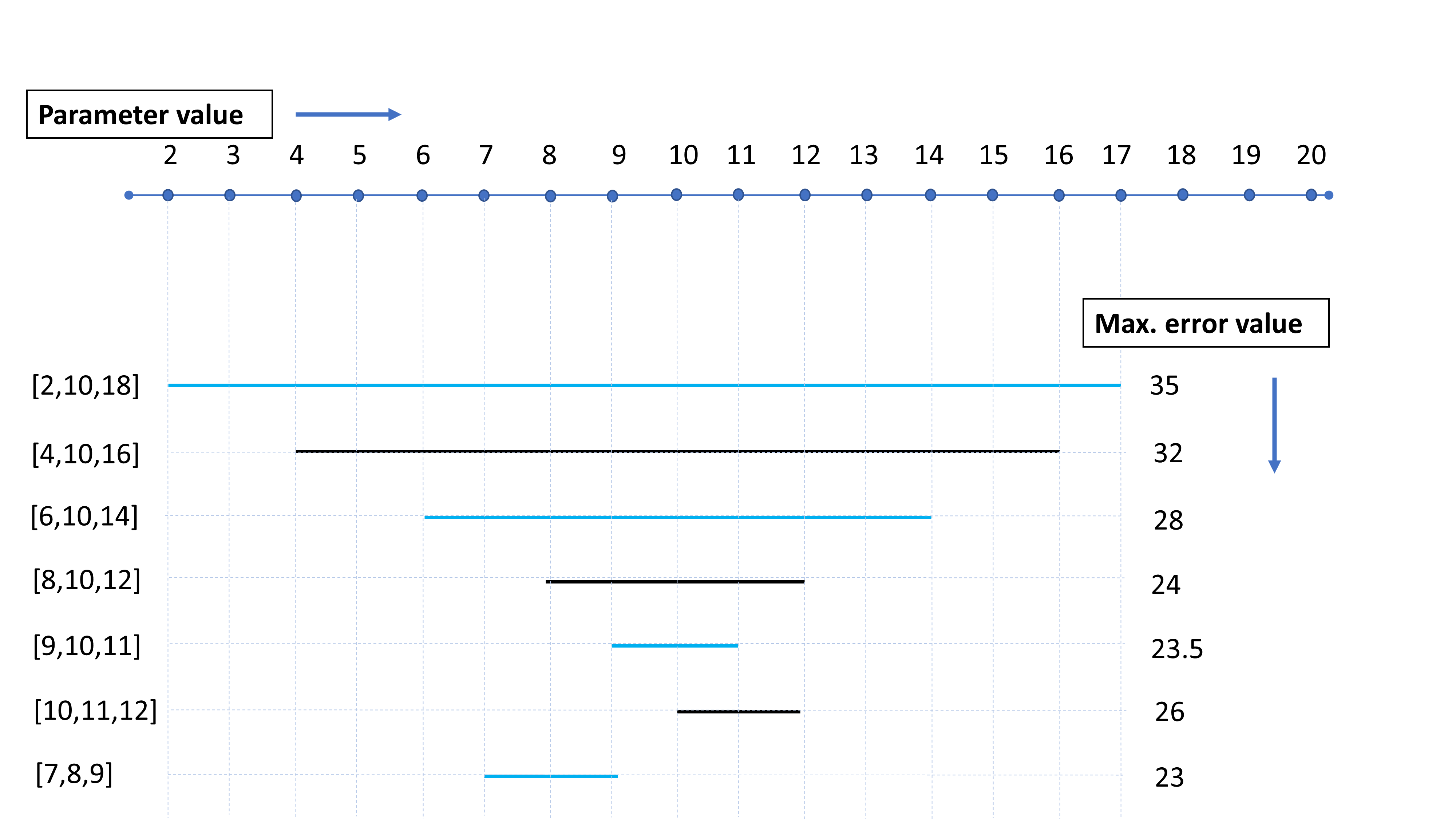}
		} 
\caption{Sensitivity analysis: The plots show that our model predicts more accurately when pre-trained on  a more suitable and narrow range of pre-training parameters. More so, the accuracy is higher for parameters on the lower side of the average node degree as Power law networks have a right skewed node degree distribution with a high number of nodes with low node degree.}
\label{fig:sensitivity}
\vspace{-4mm}
\end{figure}

We saw that the parameter $m$ is analogous to node degree as it represents number of edges attached to the new node while the artificial network is being constructed.
We conduct a sensitivity analysis on the prediction performance towards parameter range. This helps us understand the model behavior and select best training parameters for the given average node degree of the test network.


Mean error over the unobserved values of the Virgili Emails network, when autoencoder is pre-trained over various parameter windows, is shown in Fig. \ref{fig:sensitivity}[a]. A more comparative visual display of the results can be seen in Fig. \ref{fig:sensitivity}[b]. We can see that for an average node degree of 9.6, the autoencoder performs better when trained on narrower windows as compared to wider. Thus, the performance improves from [2,10,18] to [9,10,11]. We also observed that a narrow window placed on the lower side of the target node degree gives a slightly better performance than window placed on the higher side of the spectrum, i.e., [7,8,9] performs better than [10,11,12], even if both the windows are equally narrow. This is because our target network is Power law in nature and a lot of nodes have degrees on the lower side of the average node degree. 



\section{Conclusion}
\label{sec:conclusion}

In this paper, 
we have shown how to optimally pre-train a neural network when only sparse samples are available from the target network.
We call this model ``Oracle Search Pre-training'' (OSP) as it helps us search the optimal pre-training parameter values specifically for the given social network.
We use artificially generated data to compensate for the scarcity of real-world measurements to effectively train a neural network. The low-rankness of the distance matrices is leveraged. 

Prediction performance is evaluated on three real-world networks, namely, Virgili Emails, Facebook, and Train Bombing network, under different variations of the model, and is also compared with a state-of-the-art Low-rank Matrix Completion approach. The model infers distances within one hop of the original value even when only 1\% of measurements are sampled.
We also study sensitivity of the model towards the tuning parameter values and its relation with average node degree of the social network.
It is worth mentioning that though we illustrate results on social networks, the method can be generalized to other network domains by adapting the synthetic data.

In the future, we would like to test our model for directed networks and networks from other domains such as crime networks. The effect of other network parameters, such as clustering coefficient, on reconstruction can also be studied.






\bibliographystyle{ACM-Reference-Format}
\bibliography{sample-sigconf}









\end{document}